\documentclass[runningheads,a4paper, 10pt]{llncs}
\usepackage{amssymb}
\usepackage{graphicx}
\usepackage{url}
\usepackage{multirow}
\newcommand{\keywords}[1]{\par\addvspace\baselineskip
\noindent\keywordname\enspace\ignorespaces#1}
\usepackage{booktabs}
\usepackage{jlcode}
\begin{document}

\mainmatter
%================================================
\newcommand{\C}{\mathbb{C}}
\newcommand{\R}{\mathbb{R}}
\newcommand{\bx}{\mathbf{x}}
\newcommand{\ra}[1]{\renewcommand{\arraystretch}{#1}}

%==== FILL IN ====================================
\title{HomotopyContinuation.jl:~A package for homotopy continuation in Julia}  % Full title
\titlerunning{HomotopyContinuation.jl} % Short title
\author{Paul Breiding\inst{1} \and  Sascha Timme\inst{2}}
\authorrunning{Breiding - Timme}
\institute{
Max Planck Institute for Mathematics in the Sciences Leipzig, Germany\\
\email{breiding@mis.mpg.de},\\
\url{http://personal-homepages.mis.mpg.de/breiding/}
\and
Technische Universit\"at Berlin, Germany\\
\email{timme@math.tu-berlin.de},\\
\url{http://page.math.tu-berlin.de/~timme/}
}
\maketitle

\begin{abstract}
We present the Julia package HomotopyContinuation.jl, which provides an algorithmic framework for solving polynomial systems by numerical homotopy continuation.
We introduce the basic capabilities of the package and demonstrate the software on an illustrative example.
We motivate our choice of Julia and how its features allow us to improve upon existing software packages
with respect to usability, modularity and performance.
Furthermore, we compare the performance of HomotopyContinuation.jl to the existing packages Bertini and PHCpack.
\end{abstract}

%------------------------------------------------------------
\section{Introduction}
Numerical algebraic geometry is concerned with the study of algebraic varieties by using numerical methods.
The main computational building block therein is \emph{homotopy continuation} which is a technique to
approximate zero-dimensional solution sets of polynomial systems $F: \C^n \rightarrow \C^n$.
The idea is that one first forms another polynomial system $G$ related to $F$ in a prescribed way,
which has known or easily computable solutions. Then
the systems $G$ and $F$ can be connected by setting up a homotopy $H:\C^n \times [0,1]\rightarrow \C^n$. An example for this would be the linear homotopy~$H(x, t) = (1-t) F + t G$. For a properly formed homotopy, there are
continuous solution paths leading from the solutions of $G$ to those of $F$ which may be followed
using predictor-corrector methods.
Singular solutions of~$F$ cause numerical difficulties, so singular endgames \cite{numsol} are typically employed.

There are several software packages publicly available to make computations with homotopy continuation such as Bertini \cite{bertini} and PHCpack \cite{phcpack}.
We add the new and actively developed package HomotopyContinuation.jl\footnote{\url{www.JuliaHomotopyContinuation.org}} to that list. The package is programmed in Julia \cite{julia}, which has recently gained much popularity in the numerical mathematics community. HomotopyContinuation.jl offers new and innovative features as well as a flexible design, which allows the user to adapt the code to the structure of their specific polynomial systems with little effort.
%------------------------------------------------------------
\section{Functionality}
HomotopyContinuation.jl aims at having an intuitive user interface. Assume we are interested in the solution set of the polynomial system
\begin{equation}
      F := \left[\begin{array}{c}
            x^2 + y^2 - 1 \\
            3x - 2y
      \end{array}\right]
\end{equation}
which is the intersection of a quadric with a line. The code to solve this system is as follows:
\begin{lstlisting}
using HomotopyContinuation # load package

@polyvar x y # we define variables x and y
solve([x^2+y^2-1, 3x-2y]) # define F and solve the system 
\end{lstlisting}
In the background the software first constructs the total degree start system
\begin{equation}\label{start}
  G := \left[x^2-1 \atop y-1\right]
\end{equation}
and then defines the homotopy $H(x,t):=(1-t)F+\gamma t G$ where $\gamma \in \C$
is choosen randomly. The two solutions $(-1, 1)$ and $(1, 1)$ of $G$ are tracked towards the solutions of $F$. By default, we use the classical Runge-Kutta predictor and Newton's method for correction.
Internally all computations are executed in the complex projective plane $\mathbb{P}^2$ on a (local) affine coordinate patch. In general, envoking the \texttt{solve()} command on any square system of polynomials will let HomotopyContinuation.jl generate a total degree starting system like (\ref{start}).

% This is not all implemented so far but will be in July
HomotopyContinuation.jl also features a predictor-corrector scheme for overdetermined systems of polynomials $F:\C^N \rightarrow \C^n$ with $N < n$. However, in the overdetermined case there is no way to automatically generate a suitable starting system, but the user has to provide it. Furthermore, the input to HomotopyContinuation.jl is not limited to explicitly defined polynomial systems. Custom-defined homotopies are allowed.
An example for a custom-defined homotopy for a family of overdetermined systems is given in Section \ref{sec:symmetroid}.

In order to deal with singular solutions, an endgame strategy which combines the power series \cite{numsol,polyhedral:endgame} and Cauchy endgame \cite{numsol} is implemented.
The solution can be computed in serial-processing as well as in parallel on a single machine by multiple threads.

\section{Technical contribution}
Existing software packages are, as most scientifc software, written in a fast, statically compiled language like C or C\texttt{++}.
They then have to rely on files as input and output format, which can be cumbersome to write and parse, or they
build a wrapper in a dynamic language like Python to allow the user to interact with the core software.
While such a wrapper is preferable to a file based user interface it also has disadvantages. It puts an additional development
and maintenance burden on the software authors and ultimately limits the flexibility of possible user input.

By contrast, HomotopyContinuation.jl is completely written in Julia, a high-level, dynamic programming language.
There is no separation between the computational core and a wrapper with which the user interacts, everything is pure Julia.
Julia programs are organized around \emph{multiple dispatch}, which allows built-in and user-defined functions to
be overloaded for different combinations of argument types.
With its modular design HomotopyContinuation.jl exploits Julia's architecture. It is easy for users to extend and modify the
capabilities of the package and to adapt the program to specific applications.
An illustration of this is given in the following section, where we explain how to use the modular design for creating a homotopy that computes singular points on \emph{symmetroids}.

Julia's LLVM-based just-in-time (JIT) compiler combined with the language's design allows to approach and often match the performance of C.
For specific applications one can even surpass the performance of conventional C programs by making use of Julia's
metaprogramming capabilities and its JIT compiler. One of these specific applications, which is of particular interest in the context of homotopy continuation, is the evaluation of polynomials.
Let $f$ be a polynomial with support $A\subset \mathbb{N}^n$. Generating optimal source code to evaluate
polynomials with support $A$ moves work from runtime to compile time, a tradeoff well worth if the same polynomials are evaluated very
often, as it is the case during homotopy continuation. Horner's method for polynomials over the reals
or a Goertzel-like method for complex polynomials \cite[Section 4.6.2]{Knuth:1997:ACP:270146} may be employed to reduce the number of operations.
Processor instructions like fused multiply-add (FMA) improve the performance and numerical accuracy.
  An experimental implementation of this idea by the second author is available under \url{https://github.com/JuliaAlgebra/StaticPolynomials.jl}. It also possible
  to use this optionally with HomotopyContinuation.jl.

%------------------------------------------------------------

\subsection{Implementing custom homotopies -- an example}\label{sec:symmetroid}
Above, we emphasized the \emph{modular design} of HomotopyContinuation.jl and claimed that it is useful for
creating homotopies for specific problems.
A generic homotopy like the straight-line homotopy
built from the total degree starting system (\ref{start}) is not suited for highly structured problems.
In fact, treating structured problems with structured homotopies may be decisive in making a computation feasible.
The following example illustrates this.

Let $\mathcal{A}=(A_0,A_1,A_2,A_3) \in \textrm{Sym}(\R^{n\times n})^{\times 4}$ be a 4-tuple of real symmetric matrices. The associated \emph{symmetroid} $S_\mathcal{A}$ is the hypersurface in complex projective 3-space $\mathbb{P}^3$ given by the polynomial
$$f_\mathcal{A}(x_0, x_1, x_2, x_3) := \det(x_0 A_0 + x_1 A_1 + x_2 A_2 + x_3 A_3).$$
Already studied by Cayley \cite{cayley}, symmetroids are objects of interest at the intersection between algebraic geometry and optimization. Let us explain the connection to the latter. For a point $x=(x_0,\ldots,x_3)\in \mathbb{P}^3_\mathbb{R}$ with $x_0\neq 0$ let us write $z_i = \frac{x_i}{x_0}$ for affine coordinates. The set of real points $x=(x_0,\ldots,x_3)\in \mathbb{P}^3_\mathbb{R}$ such that $ A_0 + z_1 A_1 + z_2 A_2 + z_3 A_3$ is positive semi-definite is called a \emph{spectrahedron}~\cite{vinzant} and we denote it by $\Sigma_\mathcal{A}$. Spectrahedra are feasible sets in semi-definite programming, which is a generalization of linear programming \cite{semi2,semi1}. For instance, problems as finding the smallest eigenvalue of a symmetric matrix or optimizing a polynomial function on the sphere can be formulated as a semidefinite programme. Because a linear function on a spectrahedron attains its maximum in a real singular point of the boundary with a positive probability, the number of singularities on the boundary of $\Sigma_\mathcal{A}$ matters. The boundary of $\Sigma_\mathcal{A}$ is~$\Sigma_\mathcal{A}\cap S_\mathcal{A}$.

If $A_1, A_2, A_3, A_4$ are generic, the singular locus of the symmetroid $S_{\mathcal{A}}$
consists of ${n+1\choose 3}$ isolated points. It is known how to construct a tuple $\mathcal{B}=(B_0,B_1,B_2,B_3)$ together with all of the associated ${n+1\choose 3}$ singular points on~$S_{\mathcal{B}}$. Moreover, the construction is such that the ${n+1\choose 3}$ singular points of $S_{\mathcal{B}}$ are all real; see, e.g., \cite[Theorem 1.1]{sanyal}. By contrast, a tuple $\mathcal{B}=(B_0,B_1,B_2,B_3)$ with $\Sigma_\mathcal{B}\cap S_\mathcal{B} = S_\mathcal{B}$, i.e., a tuple $\mathcal{B}$ for which all the associated singular points are at the same time points on the spectrahedron is only known for $n=4$. This is due to work by Degtyarev and Itenberg \cite{quartic}. In \cite{sturmfels} Sturmfels poses the question:
\begin{quote}
\emph{How many of the  ${n+1\choose 3}$ singular points of $S_\mathcal{A}$ can lie on the boundary of~$\Sigma_\mathcal{A}$?}
\end{quote}

By using homotopy continuation we can compute all the ${n+1 \choose 3}$ singular points on a symmetroid $S_\mathcal{A}$, from which we can check how many of them actually lie on the boundary of the spectrahedron. This way we advance in answering Sturmfels' question.
We are currently working on a full featured implementation of the symmetroid-homotopy and will publish it in the near future. For the rest of this subsection let us explain the idea and sketch how an implementation of a symmetroid-homotopy in HomotopyContinuation.jl could look like.

To study the singularities of~$S_\mathcal{A}$ we are interested in the zeros of the system
\begin{equation}\label{det_system} F_\mathcal{A}(x_0, x_1, x_2, x_3) := \left(f_\mathcal{A},
      \frac{\partial f_\mathcal{A}}{ \partial x_0},
      \frac{\partial f_\mathcal{A}}{ \partial x_1},
      \frac{\partial f_\mathcal{A}}{ \partial x_2},
      \frac{\partial f_\mathcal{A}}{ \partial x_3}
      \right) \;.\end{equation}
A homotopy from a symmetroid $S_\mathcal{B}$ to a symmetroid $S_\mathcal{A}$ is then defined as
\begin{equation}\label{eq:symmetroid_homotopy}
  H_{\mathcal{A}, \mathcal{B}}(x,t) := F_{(1-t) \mathcal{A} + t\mathcal{B}}(x) \,.
\end{equation}
Note that the number of monomials in $H_{\mathcal{A}, \mathcal{B}}(x,t)$ in $(x_0,x_1,x_2,x_3,t)$ for the generic choice of symmetric matrices is
$(n+1){n+3 \choose n} + 4(n+1){n+2 \choose n-1}$. For $n=20$ this number is 166551. The size of the polynomials prevents us from working with explicit expressions in the monomial basis. Already evaluating $F_\mathcal{A}$ and its Jacobian by considering the representation of $f_\mathcal{A}$ in the monomial basis
becomes prohibitively expensive. On the other hand, the number of solutions of the system~(\ref{det_system}) is ${21 \choose 3} = 1330$, which is reasonably small.

Nevertheless, homotopy continuation algorithms never require to have the polynomial written down explicitly.  What is needed for tracking the solution paths of a homotopy $H(x,t)$ is a function to evaluate $H(x,t)$ for all $x$ and $t$ and functions for evaluating the derivatives $\frac{\partial H(x,t)}{\partial x}$ and $\frac{\partial H(x,t)}{\partial t}$. Using matrix calculus and
linear algebra, we find that the evaluation of $H_{\mathcal{A}, \mathcal{B}}$ and its Jacobian matrix at $x$ are given by the first and second order derivatives of $f_\mathcal{A}$ at $x$.
Denoting $A(x) := x_0 A_0 + x_1 A_1 + x_2 A_2 + x_3 A_3$ and $P_i(x) := A(x)^{-1}A_i$ they can be written in the following compact form:
$$
\begin{array}{rl}
      \frac{\partial f_\mathcal{A}}{ \partial x_i}(x)  &=  \det(A(x)) \textrm{tr}(P_i(x)) \\[1em]
      \frac{\partial^2 f_\mathcal{A}}{ \partial x_i \partial x_j} (x) &=
      \det(A(x)) \textrm{tr}(P_i(x))

       \textrm{tr}(P_j(x)) - \det(A(x)) \textrm{tr}(P_i(x) P_j(x))
\end{array}$$
where we used the fact that $\frac{\partial A(x)^{-1}}{\partial x_i} = -A(x)^{-1}A_i A(x)^{-1}$. The derivative of $H(x,t)$ with respect to $t$ is obtained by a similar computation. Hence, the evaluation of $F_\mathcal{A}$ and its partial derivative can be done efficiently, because evaluating determinants can be done efficiently.

% A homotopy from a symmetroid $S_\mathcal{B}$ to a symmetroid $S_{A}$ is then defined as
% \begin{equation}\label{eq:symmetroid_homotopy}
%   H_{\mathcal{A}, \mathcal{B}}(x,t) := F_{(1-t) \mathcal{A} + t\mathcal{B}}(x) \,.
% \end{equation}
We use the aforementioned construction from \cite[Theorem 1.1]{sanyal} for building a start system $F_\mathcal{B}$. The Runge-Kutta predictor scheme and the overdetermined Newton corrector are employed for tracking the solutions from $F_\mathcal{B}$ to~$F_\mathcal{A}$.

Implementing this homotopy in existing software packages is very onerous and slow since the predefined interfaces can only handle the
polynomial representation of $H_{\mathcal{A}, \mathcal{B}}$.
By contrast, in HomotopyContinuation.jl the homotopy can be implemented in an efficient way. Since everything is defined in Julia,
we have a full-fledged programming language at our hand to evaluate $H_{\mathcal{A}, \mathcal{B}}$.
An illustrative example of the subset of the code necessary to handle $H_{\mathcal{A}, \mathcal{B}}$ in HomotopyContinuation.jl is depicted in Figure
\ref{fig:symmetroid_code}.

\begin{figure}
\begin{lstlisting}
import HomotopyContinuation.Homotopies:
  AbstractHomotopy, evaluate!, jacobian!

struct SymmetroidHomotopy{T} <: AbstractHomotopy
  A_tuple::NTuple{4, Symmetric{T, Matrix{T}}}
  B_tuple::NTuple{4, Symmetric{T, Matrix{T}}}
end

# The SymmetroidHomotopy is a homotopy of 5 polynomials
# in 4 variables
Base.size(::SymmetroidHomotopy) = (5, 4)

# This computes H(x,t) and stores the result in u
function evaluate!(u, H::SymmetroidHomotopy, x, t, cache)
  tuple_at_t = (1-t) .* H.A_tuple .+ t .* H.B_tuple
  A = sum(x[i] * tuple_at_t[i] for i in 1:4)
  # Compute the inverse and determinant of the matrix A
  inv_A, det_A = inv(A), det(A)
  u[1] = det_A
  for i=1:4
    u[i+1] = det_A + trace(inv_A * tuple_at_t[i])
  end
end

# This computes the Jacobian of H at (x,t) and stores
# the result in U
function jacobian!(U, H::SymmetroidHomotopy, x, t, cache)
  tuple_at_t = (1-t) .* H.A_tuple .+ t .* H.B_tuple
  A = sum(x[i] * tuple_at_t[i] for i in 1:4)
  inv_A, det_A = inv(A), det(A)
  P = [inv_A * tuple_at_t[i] for i in 1:4]
  traces = trace.(P)
  U[1,:] = det_A .* traces
  for i=1:4, j=i:4
    U[1+i,j] = U[1+j,i] = det_A * traces[i] * traces[j] -
                          det_A * trace(P[i] * P[j])
  end
end
\end{lstlisting}
\caption{Subset of the code necessary to track solutions of the homotopy $H_{\mathcal{A}, \mathcal{B}}$.
  In addition it is necessary to define a function \texttt{dt!}, which evaluates $\frac{\partial}{\partial t}H(x,t)$.
  Furthermore, it is possible to
  define a function \texttt{evaluate\_and\_jacobian!} that evaluates $H(x,t)$ and computes its Jacobian simultaneously.
  This is in particular useful here due to the shared structure of the derivatives.
 Although this code is able to solve the problem, it is written in an illustrative style.
  In a full featured implementation we would define an additional \texttt{cache} object to precallocate structures to avoid unnecessary temporary allocations.}\label{fig:symmetroid_code}
\end{figure}

%------------------------------------------------------------
\section{Comparison}

We compare HomotopyContinuation.jl against the established software packages Bertini and PHCpack.
For this we pick a range of real-world polynomial systems of different type, presented in Table \ref{table:systems}\footnote{The authors discovered the examples in the excellent database of Jan Verschelde available at \url{http://homepages.math.uic.edu/~jan/}},
and solve each polynomial system 10 times.

\begin{table*}\centering
  \caption{Overview of the polynomial systems choosen for the comparison. In the characteristics $n$ is the number
  of unknowns, $D$ is the B\'ezout number of the system and MV is the mixed volume. The system were taken
  from the database by Jan Verschelde.}\label{table:systems}
  \begin{tabular}{@{}llrrcrrrrr@{}}\toprule
  \multicolumn{3}{c}{Polynomial systems} & \phantom{abc}& \multicolumn{3}{c}{Characteristics}
    & \phantom{abc} & \multicolumn{2}{c}{\# Roots} \\
  \cmidrule{1-3}  \cmidrule{5-7} \cmidrule{9-10}
  Name & Description & Ref && $\;n\;$ & $D$ & MV && $\C$ & $\R$ \\ \midrule
  cyclic7 & The cyclic 7-roots problems &\cite{cyclic} && $7$ & 5,024 & $924$ && 924 & 56\\
  ipp2 &  The 6R inverse position problem &\cite{ipp2} && $11$ & 1,024 & $288$ && 16 & 0\\
  heart & The heart-dipole problem & \cite{heart}  && $8$ & 576 & $121$ && 4 & 2\\
  katsura11 & A problem of magnetism in physics &\cite{katsura} && 12 & 2,048 & 2,048 && 2,048 & 326\\
  \bottomrule
  \end{tabular} \linebreak
\end{table*}

In particular, we take the perspective of a non-expert user and solve every system without any modification to the default parameters of the respective
software packages.
The only excemption is that for Bertini we distinguish between a version which uses adaptive precision \cite{adaptive}
and one which uses standard 64 bit floating
point arithmetic since HomotopyContinuation.jl as well as PHCpack also only compute by default with standard 64 bit floating
point arithmetic.

We compare the packages with respect to outside observation. This is the number of times the correct number of
solutions, the average number of solutions found, the average and median number of reported path failures
(since these introduce uncertainity about the correctness of the result) and the average runtime.
The results of the comparison are presented in Table \ref{table:results}.
They were run on a MacBook Pro with a 2 GHz Intel i5-6360U CPU. We used MacOS 10.13.4 and Julia 0.6.2,
Bertini v1.5.1 and PHCpack v2.4.52 and HomotopyContinuation.jl v0.2.0-alpha.2.

\begin{table}\centering
    \caption{The results obtained for the systems in Table \ref{table:systems} using serial processing.
    }\label{table:results}

    \begin{tabular}{@{}cclcrrrrrrrrr@{}}\toprule
       &&  && \multicolumn{2}{c}{\# Solutions} & \phantom{a} &\multicolumn{2}{c}{\# Failed paths} & \phantom{a} & Runtime \\
      \cmidrule{5-6} \cmidrule{8-9} \cmidrule{11-11}
      Systems && Package && \footnotesize{correct} & avg. & \phantom{a} & avg. & med. & \phantom{a} & \multicolumn{1}{c}{avg.} \\ \midrule
      \multirow{4}{*}{cyclic7}
      && Bertini && 8/10 \hspace{0.1em} & 923.3 && 1196.5 & 1300 && 48.93s\\
      && Bertini (adaptive precision) && 10/10 \hspace{0.1em} & 924.0 && 0 & 0 && 1028.21s\\
      && PHCpack && 0/10 \hspace{0.1em} & 918.4 && 5.6 & 5.5 &&6.48s\\
      && HomotopyContinuation.jl && 10/10 \hspace{0.1em} & 924.0 && 0 & 0 && 8.38s\\[0.5em]
      \multirow{4}{*}{heart}
      && Bertini && 10/10 \hspace{0.1em} & 4.0 && 66.0 & 73.5 && 4.88s\\
      && Bertini (adaptive precision) && 10/10 \hspace{0.1em} & 4.0 && 0 & 0 && 30.63s\\
      && PHCpack && 10/10 \hspace{0.1em} & 4.0 && 16.5 & 16 &&1.33s\\
      && HomotopyContinuation.jl && 10/10 \hspace{0.1em} & 4.0 && 0 & 0 && 1.39s\\[0.5em]
      \multirow{4}{*}{ipp2}
      && Bertini && 10/10 \hspace{0.1em} & 16.0 && 0.5 & 0 && 10.03s\\
      && Bertini (adaptive precision) && 10/10 \hspace{0.1em} & 16.0 && 0 & 0 && 13.15s\\
      && PHCpack && 10/10 \hspace{0.1em} & 16.0 && 272 & 272 &&6.67s\\
      && HomotopyContinuation.jl && 10/10 \hspace{0.1em} & 16.0 && 0 & 0 && 3.07s\\[0.5em]
      \multirow{4}{*}{katsura11}
      && Bertini && 8/10 \hspace{0.1em} & 2047.7 && 0.2 & 0 && 28.97s\\
      && Bertini (adaptive precision) && 10/10 \hspace{0.1em} & 2048.0 && 0 & 0 && 28.88s\\
      && PHCpack && 0/10 \hspace{0.1em} & 2043.7 && 2.3 & 2.0 &&179.13s\\
      && HomotopyContinuation.jl && 10/10 \hspace{0.1em} & 2048.0 && 0 & 0 && 9.30s\\[0.5em]
  \bottomrule
  \end{tabular}
\end{table}

%------------------------------------------------------------


\begin{thebibliography}{4}

\bibitem{semi2}
Alizadeh, F. (1995).
\newblock Interior point methods in semidefinite programming with applications to combinatorial optimization.
\newblock SIAM Journal on Optimization, 5(1):13-51.

\bibitem{adaptive}
  Bates, D. J., Hauenstein, J. D., Sommese, A. J., \& Wampler, C. W. (2008).
  \newblock Adaptive multiprecision path tracking.
  \newblock SIAM Journal on Numerical Analysis, 46(2), 722-746.

\bibitem{bertini}
      Daniel J. Bates, Jonathan D. Hauenstein, Andrew J Sommese,
      and Charles W. Wampler.
      \newblock Bertini: Software for Numerical Algebraic Geometry.
      \newblock Available at bertini.nd.edu with permanent doi: dx.doi.org/10.7274/R0H41PB5.
\bibitem{julia}
Bezanson, J., Edelman, A., Karpinski, S., \& Shah, V. B. (2017).
\newblock Julia: A fresh approach to numerical computing.
\newblock SIAM review, 59(1), 65-98.


\bibitem{cyclic}
Bj\"orck, G., \& Fr\"oberg, R. (1991).
\newblock A faster way to count the solutions of inhomogeneous systems of algebraic equations, with applications to cyclic n-roots.
\newblock Journal of Symbolic Computation, 12(3), 329-336.


\bibitem{cayley}
Cayley, A. (1869/71).
\newblock A memoir on quartic surfaces.
\newblock Proc. London Math. Soc. 3  19-69. [Collected Papers, VII, 133-181; see also the sequels on pages 256-260, 264-297].


\bibitem{quartic}
Degtyarev, A., Itenberg,  I. (2011).
\newblock On real determinantal quartics.
\newblock Proceedings of the G\"okova Geometry Topology Conference 2010.

\bibitem{polyhedral:endgame}
Huber, B., \& Verschelde, J. (1998).
\newblock Polyhedral end games for polynomial continuation.
\newblock Numerical Algorithms, 18(1), 91-108.

\bibitem{katsura}
Katsura, S. (1990).
\newblock Spin glass problem by the method of integral equation of the effective field.
\newblock New Trends in Magnetism, 110-121.

\bibitem{Knuth:1997:ACP:270146}
Knuth, D. E (1997).
\newblock The art of computer programming, volume 2 (3rd ed.): Seminumerical Algorithms.
\newblock Addison-Wesley Longman Publishing Co.

\bibitem{heart}
Nelson, C. V., \& Hodgkin, B. C. (1981).
\newblock Determination of magnitudes, directions, and locations of two independent dipoles in a circular conducting region from boundary potential measurements.
\newblock IEEE Transactions on Biomedical Engineering, (12), 817-823.

\bibitem{semi1}
Ramana, M., Goldman, A. J. (1995).
\newblock Some geometric results in semidefinite programming.
\newblock Jnl. Glob. Opt, 7:33-50.

\bibitem{sanyal}
Sanyal, R. (2011).
\newblock On the derivative cones of polyhedral cones.
\newblock Advances in Geometry, Volume 13, Issue 2, Pages 315-321

\bibitem{sturmfels}
Sturmfels, B. (2014).
\newblock Spectrahedra and their shadows.
\newblock Talk at the Simons Institute Workshop on
Semidefinite Optimization, Approximation and Applications. Slides available at \url{https://simons.berkeley.edu/sites/default/files/docs/2039/slidessturmfels.pdf}.

\bibitem{phcpack}
Verschelde, J. (1999).
\newblock Algorithm 795: PHCpack: A general-purpose solver for polynomial systems by homotopy continuation.
\newblock ACM Transactions on Mathematical Software (TOMS), 25(2), 251-276.

\bibitem{vinzant}
Vinzant, C. (2014).
\newblock What is ... a Spectrahedron?
\newblock Notices of the AMS, 61(5).


\bibitem{ipp2}
Wampler, C., \& Morgan, A. (1991).
\newblock Solving the 6R inverse position problem using a generic-case solution methodology.
\newblock Mechanism and Machine Theory, 26(1), 91-106.

\bibitem{numsol}
Wampler, I. C. W. (2005).
\newblock The numerical solution of systems of polynomials arising in engineering and science.
\newblock World Scientific.





\end{thebibliography}
\end{document}